\documentstyle[seceq,epsf,twoside]{ptptex}
\notypesetlogo  %comment in if to eliminate PTPTeX logo
\title{Ground State Baryons in Covariant Level-Classification Scheme}
\author{%
Muneyuki {\sc Ishida}$^1$ and Shin {\sc Ishida}$^2$  }
\inst{%
$^1$ Department of Physics, Meisei University, 
Hino 191-8506, Japan\\
$^2$Research Institute of Quantum Science, 
College of Science and Technology\\
Nihon University, Tokyo 101-0062, Japan
}
\abst{%
In the covariant level-classification scheme of hadrons based on 
$\tilde U(12)_{SF}\bigotimes O(3,1)_L$ symmetry, 
the ground state baryons and antibaryons are assigned 
to the $({\mib 12\times 12\times 12})_{\rm Sym}={\bf 364}$-multiplet in
the spin-flavor $\tilde U(12)_{SF}$ symmetry. 
This multiplet includes, in addition to the ordinary ${\bf 56}$-multiplet 
of the static $SU(6)_{SF}$, the extra chiral ${\bf 56}^\prime$ and  
${\mib 70}$-multiplets, called ``chiralons,"
which are expected to exist with masses in 1 $\sim$ 2 GeV region.
Their electro-magnetic properties, magnetic moments and radiative decay amplitudes,
are investigated.
A longstanding puzzle,
that the predicted value of the width $\Gamma (\Delta\rightarrow N\gamma)$ by the conventional treatment
is inconsistently small with the experimental one, 
 may be solved by the mixing effect of ${\bf 56}$-states with 
the chiral ${\bf 56}^\prime$-states.
}
\begin{document}
\maketitle

\setcounter{tocdepth}{4}

\section{Introduction}

In the covariant level classification scheme of hadrons presented by one of the authors 
S. I.(, referred to as II\cite{S.I.}), 
the level spectra of hadrons including light quarks in lower mass region are classified 
as the representations of $\tilde U(12)_{SF}\bigotimes O(3,1)_L$ group\cite{U12}, which is a 
covariant generalization of $SU(6)_{SF}\bigotimes O(3)_L$ in the non-relativisitic quark model (NRQM).
The light quark baryons and anti-baryons in the ground state are assigned as
the $({\mib 12}\times {\mib 12}\times {\mib 12})_{\rm Sym}={\bf 364}$-multiplet 
in the spin-flavor $\tilde U(12)_{SF}$ group.
This multiplet includes, in addition to the ordinary ${\bf 56}$ 
of the static $SU(6)_{SF}$, the extra ${\bf 56}^\prime$ and  ${\mib 70}$, 
which are predicted to exist with masses in 1 $\sim$ 2 GeV region 
and are out of the conventional framework.
The essential reason for obtaining the extra multiplets is the inclusion of 
$\rho$-spin degree of freedom for the constituent quarks
in addition to the ordinary Pauli- ($\sigma$-) spin, which correspond to the 
$\rho\bigotimes\sigma$ decomposition of Dirac $\gamma$-matrices.
The $\rho$-spin freedom is necessary for covariant description of composite hadrons.
The wave functions (WFs) of non-relativistic {\bf 56} states 
correspond to the boosted Pauli-spinors, the multi-Dirac spinors consisting of Dirac spinors with positive
$\rho_3$ spin. 
On the other hand the WFs of the extra ${\bf 56}^\prime$ and  ${\mib 70}$ states include 
the multi-Dirac spinors consisting of 
negative $\rho_3$-spin Dirac spinors, which is related to the ones with the positive  $\rho_3$-component 
through chiral transformation. These extra states are called chiral states (``chiralons").\cite{titi} 

In this talk we investigate the electromagnetic properties of ground state baryons, their magnetic moments
and radiative decays. For this purpose, at first,
we construct their covariant spinor-flavor WFs. 

%There is a longstanding problem in the radiative decay  $\Delta (1232)\rightarrow N\gamma$:
%The decay width predicted by NRQM is small, and inconsistent with the experimental data.
%We will show this problem may be solved by considering the mixing between the  
%non-relativistic {\bf 56}-plet and relativistic ${\bf 56}^\prime$-plet.

\section{Flavor-Spinor WF of Ground State Baryons}

\subsection{Spinor WF and chiral states} 

In order to make clear the physical background for the chiral states we describe the
spinor WF of the ground baryons, neglecting the internal space-time variables.
First we define the Dirac spinor for quarks $W_q$ 
as BW spinors with single index by   
\begin{eqnarray}
{\rm Dirac\ spinor}\ \ \ 
\psi_{q,\alpha} (X) &=& \sum_{{\bf P},r} [ e^{iPX} W_{q,\alpha}^{(+)}(P) 
   + e^{-iPX} W_{q,\alpha}^{(-)}(P) ]  \nonumber\\
   & &W_{q,\alpha}^{(+)}(P) = u_\alpha (P),
       \ \ \ W_{q,\alpha}^{(-)}(P) = u_\alpha (-P).  \ \ \ \ \ \ 
\label{eq15}
\end{eqnarray}
They take the following form at the hadron rest frame as
\begin{eqnarray}
 W_{q,r}^{(+)} ({\mib P}={\mib 0}) &=& \left( \begin{array}{c} \chi_r \\ 0 \end{array} \right)
 \ \ \rho_3=+\ ,\ \ \ \ 
 W_{q,r}^{(-)}({\mib P}={\mib 0}) = \left( \begin{array}{c} 0\\  \chi_r  \end{array} \right)
\ \  \rho_3=-\ .\ \ \ \ \ 
\label{eq17}
\end{eqnarray}
It is to be noted that all Dirac spinors with positive and negative values of 
$\rho_3$ spin for quarks are required as members of complete set of expansion bases 
inside of hadrons.
The spinor WF for ground states of baryons are given by
tri-Dirac spinors as BW spinors with three indices as
\begin{eqnarray}
{\rm Baryon\ spinor}\ \ \ \ & &
  \Phi_{\alpha\beta\gamma}^{(B)} (P)\ =\  W_{q,\alpha}(P) \  W_{q,\beta}(P)\ W_{q,\gamma}(P)   \nonumber\\
   \Phi_{\alpha\beta\gamma}({\mib P}={\mib 0}); &\ \  & (\rho_3^{(1)},\rho_3^{(2)},\rho_3^{(3)}) =  \ 
    (\ +, \ +, \ + \ ):\ {\rm boosted\ Pauli\ states}\ \ \ \ \ \ \ \ \ \ \   
\label{eq19}\\
    & & \ \ \ \ \ \ \ \ (\ +, \ +,\ -\  ),\ (\ +,\ -,\ -\ ):\ {\rm ``Chiral\ States"}\nonumber
\end{eqnarray}
The WF with $(\rho_3^{(1)},\rho_3^{(2)},\rho_3^{(3)})=(+,+,+)$ are the 
multi-boosted Pauli spinors, which 
reduce to the multi-NR Pauli spinors at the rest frame, while  
the WF with the other values of $(\rho_3^{(1)},\rho_3^{(2)},\rho_3^{(3)})$
describe the chiral states, which newly appear in the covariant classification scheme. 

In our scheme hadrons are generally classified as the members of multiplet in the
$\tilde U_{SF}(12)\times O(3,1)$ scheme. The light-quark ground state baryons are 
assigned to the representations $(\underline{\bf 12}\times\underline{\bf 12}\times 
\underline{\bf 12} )_{\rm Symm}=\underline{\bf 364}$ of the $\tilde U(12)_{SF}$ symmetry.
(See, Table \ref{tabX}.) The numbers of freedom of spin-flavor WF in NRQM are  
$({\bf 6}\times {\bf 6}\times {\bf 6})_{\rm Symm.}
=\underline{\bf 56}$ for baryons and \underline{\bf 56}$^*$ for antibaryons: These numbers in COQM
are extended to $\underline{\bf 364}=\underline{\bf 182}$ (for baryons)
$+\underline{\bf 182}$ (for anti-baryons).

\begin{table}
\begin{center}
\begin{tabular}{lr@{}l}
Baryons: &   $({\bf 12} \times {\bf 12} \times {\bf 12})_{Sym}$   & $={\bf 364}={\bf 182}_B+{\bf 182}_{\bar B}$ \\
       &   {\bf 182}---- & $\left[ \begin{array}{l}   
  \begin{array}{cccc}  {\bf 56} & \ \  &  \Delta_{3/2}^{\bigcirc\hspace{-0.2cm}+}  & N_{1/2}^{\bigcirc\hspace{-0.2cm}+} \\  \end{array}\\
  \begin{array}{|ccccc|} \hline  {\bf 70} & \ \ & \Delta_{1/2}^{\bigcirc\hspace{-0.2cm}+(\bigcirc\hspace{-0.2cm}-)}
           & N_{3/2}^{\bigcirc\hspace{-0.2cm}+(\bigcirc\hspace{-0.2cm}-)}  &  N_{1/2}^{\bigcirc\hspace{-0.2cm}+(\bigcirc\hspace{-0.2cm}-)}     \\  
 {\bf 56}^\prime & \ \ & \Delta_{3/2}^{\bigcirc\hspace{-0.2cm}-(\bigcirc\hspace{-0.2cm}+)}
           & N_{1/2}^{\bigcirc\hspace{-0.2cm}-(\bigcirc\hspace{-0.2cm}+)}  &       \\  
  \hline   \end{array}\\
\end{array}\right.$    \\
  &  &   \ \ \ \ \ \ \ \ \ \ \ \ \hspace{1cm} {\bf Chiral States}\\
\end{tabular}
\end{center}
\caption{Quantum numbers of ground-state baryon multiplet in $\tilde U(12)_{SF}$ symmetry}
\label{tabX}
\end{table}

\subsection{Flavor-Spin Decomposition of WF}

The total baryon WF should be full-symmetric (except 
for the color freedom) under exchange of constituent quarks: 
The WF is denoted as $\Phi_{ABC}$, where $A=(\alpha ,a)$ etc. and $\alpha$ ($a$) represents spinor (flavor) index.
It is obtained as a product of the sub-space $\rho$, $\sigma$ and $F$(flavor) WF 
with respective symmetric properties: 
$\Phi_{ABC}=\langle_{ABC} | F\rho\sigma \rangle$. 
There are three ways to decompose the total symmetric WF, $| F_1 F_2 \rangle_S$,
into the sub-space WF, $|F_1\rangle$ and $|F_2\rangle$:
\begin{eqnarray}	
 | F_1 F_2 \rangle_S & = & | F_1 \rangle_S | F_2 \rangle_S\ , \ \ 
   | F_1 \rangle_A | F_2 \rangle_A\ ,\ \  
   \frac{1}{\sqrt{2}}(  | F_1 \rangle_\alpha | F_2 \rangle_\alpha 
                   + | F_1 \rangle_\beta  | F_2 \rangle_\beta ) \ ,
\ \ \ \ \ \ \ \ \label{eq3} 
\end{eqnarray}
where  
$|\ \ \rangle_S$, $|\ \  \rangle_{\alpha (\beta )}$ and 
$|\ \ \rangle_A$ represent the full-symmetric, $\alpha (\beta)$-type partial 
symmetric and full anti-symmetric subspace WF, respectively. 
Results are given in Table \ref{tab1}.

The $|\sigma \rangle_S$ ($|\sigma \rangle_{\alpha ,\beta}$) corresponds to 
the total spin $\frac{3}{2}(\frac{1}{2})$ WF. 
The $|F \rangle_S$ ($|F \rangle_{\alpha ,\beta}$) corresponds to the $\Delta$-decouplet ($N$-octet).
The $|F\sigma \rangle_S$ ($|F\sigma \rangle_{\alpha ,\beta}$)
corresponds to the ${\bf 56}$ ({\bf 70}) representation of 
static $SU(6)_{SF}$. 
The intrinsic parity is defined by $\hat P = \Pi_{i=1}^3 
\gamma_4^{(i)} = \Pi_{i=1}^3 \rho_3^{(i)}$. 
$|\rho ,\frac{3}{2}\rangle (=|+++\rangle)$ and $|\rho ,-\frac{1}{2}\rangle (\simeq |+--\rangle)$
have positive parity, while
$|\rho ,-\frac{3}{2}\rangle (=|---\rangle)$ and 
$|\rho ,\frac{1}{2}\rangle (\simeq |++-\rangle )$) have negative one.

\begin{table}
\begin{center}
\begin{tabular}{c|l|l}
\hline
    & positive-parity WF $|F\rho\sigma^{+\hspace{-0.25cm}\bigcirc}\rangle$
        & negative-parity WF $|F\rho\sigma^{-\hspace{-0.25cm}\bigcirc}\rangle$ \\ 
\hline
$E:{\bf 56}$ & $|\rho ,\frac{3}{2}\rangle_S |F\sigma\rangle_S$ 
             & $|\rho ,-\frac{3}{2}\rangle_S |F\sigma\rangle_S$ \\
  $\Delta_{3/2}$ & $=|\rho ,\frac{3}{2}\rangle_S |F\rangle_S|\sigma\rangle_S$
                   & $=|\rho ,-\frac{3}{2}\rangle_S |F\rangle_S|\sigma\rangle_S$ \\
  $N_{1/2}$ & \ \ \ $|\rho ,\frac{3}{2} \rangle_S \frac{1}{\sqrt{2}}
   (|F\rangle_\alpha |\sigma\rangle_\alpha + |F\rangle_\beta |\sigma\rangle_\beta )$ 
            & \ \ \ $|\rho ,-\frac{3}{2} \rangle_S \frac{1}{\sqrt{2}}
   (|F\rangle_\alpha |\sigma\rangle_\alpha + |F\rangle_\beta |\sigma\rangle_\beta )$  \\
\hline
$F:{\bf 56}^\prime$ & $|\rho ,-\frac{1}{2} \rangle_S | F\sigma \rangle_S$
                           & $|\rho ,\frac{1}{2} \rangle_S | F\sigma \rangle_S$  \\
  $\Delta_{3/2}^\prime$ & $=|\rho ,-\frac{1}{2} \rangle_S | F \rangle_S | \sigma \rangle_S$ 
                          & $=|\rho ,\frac{1}{2} \rangle_S | F \rangle_S | \sigma \rangle_S$ \\
  $N_{1/2}^\prime$ & \ \ \ $|\rho ,-\frac{1}{2} \rangle_S \frac{1}{\sqrt{2}}
   (|F\rangle_\alpha |\sigma\rangle_\alpha + |F\rangle_\beta |\sigma\rangle_\beta )$ 
                    & \ \ \ $|\rho ,\frac{1}{2} \rangle_S \frac{1}{\sqrt{2}}
   (|F\rangle_\alpha |\sigma\rangle_\alpha + |F\rangle_\beta |\sigma\rangle_\beta )$   \\ 
\hline
$G:{\bf 70}$ & $\frac{1}{\sqrt{2}}(|\rho ,-\frac{1}{2} \rangle_\alpha | F\sigma \rangle_\alpha 
                +|\rho ,-\frac{1}{2} \rangle_\beta | F\sigma \rangle_\beta )$ 
                    & $-\frac{1}{\sqrt{2}}(|\rho ,\frac{1}{2} \rangle_\alpha | F\sigma \rangle_\alpha 
                +|\rho ,\frac{1}{2} \rangle_\beta | F\sigma \rangle_\beta )$ \\ 
  $\Delta_{1/2}$ & $=| F \rangle_S \frac{1}{\sqrt{2}}
                (|\rho ,-\frac{1}{2} \rangle_\alpha | \sigma \rangle_\alpha 
                +|\rho ,-\frac{1}{2} \rangle_\beta  | \sigma \rangle_\beta )$
                   & $=-| F \rangle_S \frac{1}{\sqrt{2}}
                   (|\rho ,\frac{1}{2} \rangle_\alpha | \sigma \rangle_\alpha 
                   +|\rho ,\frac{1}{2} \rangle_\beta  | \sigma \rangle_\beta )$  \\
  $N_{3/2}$ & \ \ \ $\frac{1}{\sqrt{2}}
                (|\rho ,-\frac{1}{2} \rangle_\alpha | F \rangle_\alpha 
                +|\rho ,-\frac{1}{2} \rangle_\beta  | F \rangle_\beta ) | \sigma \rangle_S $ 
             & \ \ \ $-\frac{1}{\sqrt{2}}
                (|\rho ,\frac{1}{2} \rangle_\alpha | F \rangle_\alpha 
                +|\rho ,\frac{1}{2} \rangle_\beta  | F \rangle_\beta ) | \sigma \rangle_S $ \\
  $N_{1/2}$ & \ \ \ $\frac{1}{2}\left[ | F \rangle_\alpha (-|\rho ,-\frac{1}{2}\rangle_\alpha
                                                                       | \sigma\rangle_\alpha
                                 +|\rho ,-\frac{1}{2}\rangle_\beta | \sigma\rangle_\beta ) \right.$
              & \ \ \ $-\frac{1}{2}\left[ | F \rangle_\alpha (-|\rho ,\frac{1}{2}\rangle_\alpha
                                                                        | \sigma\rangle_\alpha
                                 +|\rho ,\frac{1}{2}\rangle_\beta | \sigma\rangle_\beta ) \right.$  \\
                     & \ \ \ \ \ \ \ $\left. +| F \rangle_\beta (|\rho ,-\frac{1}{2}\rangle_\alpha
                                                                        | \sigma\rangle_\beta
                                 +|\rho ,-\frac{1}{2}\rangle_\beta | \sigma\rangle_\alpha ) \right]$
                      & \ \ \ \ \ \ \ $\left. +| F \rangle_\beta (|\rho ,\frac{1}{2}\rangle_\alpha | \sigma\rangle_\beta
                                                                         +|\rho ,\frac{1}{2}\rangle_\beta | \sigma\rangle_\alpha ) \right]$  \\
  $\Lambda_{1/2}$ & \ \ \ $-|F\rangle_A \frac{1}{\sqrt{2}}
     (-|\rho ,-\frac{1}{2} \rangle_\alpha |\sigma\rangle_\beta
      +|\rho ,-\frac{1}{2} \rangle_\beta |\sigma\rangle_\alpha )$ 
                   & \ \ \ $|F\rangle_A \frac{1}{\sqrt{2}}
       (-|\rho ,\frac{1}{2} \rangle_\alpha |\sigma\rangle_\beta
        +|\rho ,\frac{1}{2} \rangle_\beta |\sigma\rangle_\alpha )$ \\
\hline
\end{tabular}
\end{center}
\caption{Flavor-spinor WF of ground-state $qqq$ $_{12}H_3={\bf 364}$-multiplet 
of $\tilde U(12)_{\rm SF}$ symmetry. 
}
\label{tab1}
\end{table}

As shown in Table \ref{tab1} 
${\bf 364}/2={\bf 182}$ representation is decomposed, in static $SU(6)_{SF}$, into 
${\bf 56}+{\bf 56}^\prime +{\bf 70}$ representation.
 
The $\rho$-spin WF of ${\bf 56}$ $E^{+\hspace{-0.25cm}\bigcirc}$, $|\rho ,\frac{3}{2}\rangle_S$, 
consists of 
only positive-$\rho_3$ consitituent quark spinor, and
thus, it is related directly with non-relativisitic two-component 
Pauli-spin WF in NRQM. The $E^{-\hspace{-0.25cm}\bigcirc}$ is its creation part.
It is remarkable that there appear the extra ${\bf 56}^\prime$ $F$ 
and {\bf 70} $G$ multiplets in the ground states.

The experimentally observed ${\bf 56}$-state 
(including $N(939)$ octet and $\Delta (1232)$ decouplet) is considered to be described by 
a superposition of non-relativisitc $E^{+\hspace{-0.25cm}\bigcirc}$-WF and 
relativisitic $F^{+\hspace{-0.25cm}\bigcirc}$ WF.
We expect the existence of extra ${\bf 56}$ and ${\bf 70}$ states (chiral states), 
which have the WF orthogonal to the one of ordinary {\bf 56}.  
Among the experimentally observed baryons,
$N^{+\hspace{-0.25cm}\bigcirc}(1440)$, $\Delta^{+\hspace{-0.25cm}\bigcirc} (1600)$
and $\Lambda^{-\hspace{-0.25cm}\bigcirc} (1405)$, which have inconsistently light masses with the
predictions by NRQM, are candidates for chiralons.

\subsection{WF in general frame and its conjugate}

The WF in general frame are obtained by boosting the WF at rest frame, given in Table \ref{tab1},
with the velocity $v_\mu \equiv P_\mu /M$. 
For single quark,
\begin{eqnarray}
W({\mib P}) &=& B({\mib v}) W({\mib 0}),\ \   
     \bar W({\mib P})(\equiv W({\mib P})^\dagger \gamma_4 ) 
   = \bar W({\mib 0}) \bar  B({\mib v})\ \   \nonumber\\
  B({\mib v}) &=& ch\theta + \rho_1\sigma_z sh\theta ,\ \  
  \bar B({\mib v})\equiv  \gamma_4 B({\mib v})^\dagger \gamma_4 
    = ch\theta - \rho_1\sigma_z sh\theta \ \ ({\rm for}\  {\mib v}\propto \hat{\mib z}),\nonumber\\
  ch\theta &=& \sqrt{\frac{\omega +1}{2}},\  sh\theta = \sqrt{\frac{\omega -1}{2}};\ \ 
            \omega =\frac{E}{M}, \frac{|{\mib P}|}{M}=2ch\theta sh\theta \ . 
\label{eq4}
\end{eqnarray}
In the non-relativistic limit $|{\mib v}|\rightarrow 0$,
the $ch\theta =1$ and $sh\theta =0$. The $sh\theta$ represents the relativistic recoil effect.

The baryon WF with momentum ${\mib P}$, $\Phi_{\alpha\beta\gamma}({\mib P})$, 
is obtained by operating the booster $B^{(i)}$ for respective constituents of 
$\Phi_{\alpha\beta\gamma}({\mib 0})$, as
\begin{eqnarray}
 \Phi_{\alpha\beta\gamma}({\mib P}) &=& B_\alpha^{(1)\alpha^\prime}({\mib v}) 
     B_\beta^{(2)\beta^\prime}({\mib v}) B_\gamma^{(3)\gamma^\prime}({\mib v}) 
     \Phi_{\alpha^\prime \beta^\prime \gamma^\prime }({\mib 0}).
\label{eq5}
\end{eqnarray}

In calculating the transition matrix elements of radiative decays,  
the Pauli-conjugate  $\bar\Phi (\equiv \Phi^\dagger \gamma_4^{(1)}  \gamma_4^{(2)}  \gamma_4^{(3)}  )$
is necessary to be revised, in order to give correct charge for the WF with negative $\rho_3$-spin, into
``unitary conjugate" $\bar\Phi_U$ defined by 
\begin{eqnarray}
\bar\Phi_U ({\mib v}) & \equiv &  \bar\Phi ({\mib v}) 
  (-i v\cdot\gamma^{(1)}) (-i v\cdot\gamma^{(2)})(-i v\cdot\gamma^{(3)})\ . \ \ \ \ \   
\label{R2}\\
\bar\Phi_U ({\mib v}) &=& \Phi^\dagger ({\mib v}) \Pi_{i=1}^3  \bar B^{(i)}({\mib v})\nonumber
\end{eqnarray}
$\bar \Phi_U({\mib v})$ coincides with $\Phi^\dagger$ at the rest frame ${\mib v}={\mib 0}$.
Because of $-i v\cdot\gamma^{(i)}$, $\bar\Phi_U ({\mib v})  \Phi ({\mib 0})$  
is invariant under chiral transformation $\Phi \rightarrow \Pi_{i=1}^3 e^{i\alpha\gamma_5^{(i)}} \Phi $

It should be noted that the above WF $\Phi_{\alpha\beta\gamma}({\mib v})$
are equivalent to those\footnote{The WFs  are given in the form 
$\Phi_{ABC} = \frac{\sqrt 3}{2\sqrt 2} D_{\alpha\beta\gamma ;abc} 
               + V_{\alpha\beta\gamma} \epsilon_{abc} 
              +\frac{1}{2\sqrt 6}\left( N_{\alpha [\beta\gamma],a}{}^d\epsilon_{dbc} 
                              + perm   \right)\ $,
where the decouplet WF $D_{\alpha\beta\gamma}^{abc}$ 
(the singlet WF $\epsilon_{abc}V_{[\alpha\beta\gamma ]}$ )
has completely symmetric (anti-symmetric) indices $\alpha\beta\gamma$
and $abc$ including $\Delta_{3/2}^E$, $\Delta_{3/2}^{\prime F}$ and $\Delta_{1/2}^{\prime G}$. 
While   
$\epsilon_{abd}N_{[\alpha\beta ]\gamma ,\ c}{}^d$ has partially antisymmetric
indices $\alpha\beta$ and $ab$ including $N_{1/2}^E$, $N_{1/2}^{\prime F}$ and $N_{1/2,3/2}^G$. 
}
of manifestly covariant form given in ref.\citen{SDS}.

\section{Electro-magnetic Property}

\subsection{Form of  electromagnetic interaction}

The electro-magnetic interaction of the constituent quark is given, in momentum representation, by
\begin{eqnarray}
j^{EM}_{1\mu}(P^\prime ,P) A_\mu (q) &=& 
     eQ^{(1)} \left( \frac{1}{2M}(P_\mu +P^\prime_\mu) 
+\frac{g_M}{2m_1}i\sigma_{\mu\nu}^{(1)}q_\nu \right) A_\mu (q)\ \ \ \ \ \ \ \ \ \ 
\label{eqC2}
\end{eqnarray}
where 
$ j^{EM}_{1\mu}= j^{con}_{1\mu} +j^{spin}_{1\mu} $. The first (second) term of Eq.~(\ref{eqC2}) 
comes from the convection (spin) current $j^{con}_{1\mu}$ ($j^{spin}_{1\mu}$). 
The photon $A_\mu$ is supposed to couple to the first quark denoted by index $(1)$.
By using $i\sigma_{ij}q_jA_i(q)=\sigma_k H_k(q)$ and 
$i\sigma_{4i}q_iA_4(q)+i\sigma_{i4}q_4A_i(q)= - i \rho_1\sigma_i E_i(q)$, the second term is rewritten by
\begin{eqnarray}
j^{spin}_{1\mu} A_\mu (q) &=& 
   \mu^{(1)} (\sigma_k H_k(q)  - i \rho_1^{(1)}\sigma_k^{(1)} E_k(q) ),\ \ \mu^{(1)}\equiv eQ^{(1)} \frac{g_M}{2m_1},
\label{eqC3}
\end{eqnarray}
where the $\sigma_k H_k$ term is the ordinary magnetic interaction appearing in NRQM, 
while the second term comes from the relativistic effect, where the $\rho$ and $\sigma$ spins
couple to the electric field $E_k$ directly.
It may be called ``intrinsic electric dipole" interaction.
Our effective current Eq.~(\ref{eqC2}) 
is conserved in the symmetric limit $(M=M^\prime ;\ M(M^\prime )$ being 
the initial(final) meson mass in the relevant case of the transitions between ground state baryons).

\subsection{Magnetic moment}

By using Eq.~(\ref{eqC2}) and the WF in table \ref{tab1},
we can calculate the magnetic moments ($m.m.$),
\begin{eqnarray}  
m.m. &=& \langle F\rho\sigma |  \sum_{i=1}^3  \mu^{(i)} \sigma_z^{(i)} | F\rho\sigma \rangle
  =\langle F\rho\sigma | 3 \mu^{(1)} \sigma_z^{(1)} | F\rho\sigma \rangle ,
\label{eqC4}
\end{eqnarray}
at the rest frame, supposing the baryon being polarized along $z$-direction.
The factor 3 comes from the full-symmetry of WF. 
The results are given in Tables \ref{tab8} (for octets),
\ref{tab9} (for decouplets) and \ref{tab10} (for singlet).

\begin{table}
\begin{tabular}
{|l|c@{}|@{}c|c@{}|@{}c|c@{}|@{}c@{}|@{}c|c@{}|@{}c|}
\hline
    & $\bar pp$ & $\bar nn$ & $\bar\Lambda\Lambda$ &  $\bar\Lambda\Sigma^0$ &  
  $\bar\Sigma^+\Sigma^+$ & $\bar\Sigma^-\Sigma^-$ &  $\bar\Sigma^0\Sigma^0$ &  
  $\bar\Xi^- \Xi^-$ &  $\bar\Xi^0 \Xi^0$ \\
\hline\hline 
$|N_E^{\bf 56}\rangle$ & $\frac{4\mu_u-\mu_d}{3}$ & $\frac{-\mu_u+4\mu_d}{3}$ &
 $\mu_s$ & $\frac{\mu_u-\mu_d}{\sqrt{3}}$ & $\frac{4\mu_u-\mu_s}{3}$ & 
 $\frac{4\mu_d-\mu_s}{3}$ & $\frac{2\mu_u+2\mu_d-\mu_s}{3}$ &
 $\frac{-\mu_d+4\mu_s}{3}$ & $\frac{-\mu_u+4\mu_s}{3}$ \\
\hline
Theor. &  \underline{2.793} & -1.862 & \underline{-0.613} & 1.61 & 2.69 & -1.04 & 0.825 &
               -0.507 & -1.44 \\
\hline
Exp. &  2.793 & -1.913 & -0.613(4) & 1.61(8) & 2.46(1) & -1.16(3) &  & -0.651(3) & -1.25(1) \\
\hline\hline
$|N_F^{\bf 56^\prime}\rangle$ & $\frac{4\mu_u-\mu_d}{3}$ & $\frac{-\mu_u+4\mu_d}{3}$ &
 $\mu_s$ & $\frac{\mu_u-\mu_d}{\sqrt{3}}$ & $\frac{4\mu_u-\mu_s}{3}$ & 
 $\frac{4\mu_d-\mu_s}{3}$ & $\frac{2\mu_u+2\mu_d-\mu_s}{3}$ &
 $\frac{-\mu_d+4\mu_s}{3}$ & $\frac{-\mu_u+4\mu_s}{3}$ \\
\hline
Theor. &  2.793 & -1.862 & -0.613 & 1.61 & 2.69 & -1.04 & 0.825 &
               -0.507 & -1.44 \\
\hline\hline
$|N_G^{\bf 70}\rangle$ & $\frac{2\mu_u+\mu_d}{3}$ & $\frac{\mu_u+2\mu_d}{3}$ &  
 $\frac{\mu_u+\mu_d+\mu_s}{3}$ & 0 &  $\frac{2\mu_u+\mu_s}{3}$ &  
 $\frac{2\mu_d+\mu_s}{3}$ &  $\frac{\mu_u+\mu_d+\mu_s}{3}$ &  
 $\frac{\mu_d+2\mu_s}{3}$ &  $\frac{\mu_u+2\mu_s}{3}$ \\
\hline 
Theor. &  0.931 & 0 & 0.106 & 0 & 1.037 & -0.825 & 0.106 &
               -0.719 & 0.212 \\
\hline\hline
$|N_{\frac{3}{2}G}^{\bf 70}\rangle$ & 
 ${\scriptstyle 2\mu_u+\mu_d}$ & ${\scriptstyle \mu_u+2\mu_d}$ &  
 ${\scriptstyle \mu_u+\mu_d+\mu_s}$ & ${\scriptstyle 0}$ &  ${\scriptstyle 2\mu_u+\mu_s}$ &  
 ${\scriptstyle 2\mu_d+\mu_s}$ &  ${\scriptstyle \mu_u+\mu_d+\mu_s}$ &  
 ${\scriptstyle \mu_d+2\mu_s}$ &  ${\scriptstyle \mu_u+2\mu_s}$ \\
\hline 
Theor. &  2.793 & 0 & 0.318 & 0 & 3.11 & -2.48 & 0.318 &
               -2.16 & 0.636 \\
\hline\hline
\end{tabular}
\caption{Magnetic moment of nucleon octets: Unit is the nuclear magneton $\mu_N = \frac{e}{2m_N}$.
$\mu_d=-\mu_u/2$ is assumed. The values with underlines, which are used as inputs, 
give $\mu_u=1.862 \mu_N$ and $\mu_s=-0.613 \mu_N$. 
}
\label{tab8}
\end{table}

\begin{table}
\begin{tabular}
{|l|c@{}|@{}c@{}|@{}c@{}|@{}c|c@{}|@{}c@{}|@{}c|c@{}|@{}c|c|}
\hline
    & $\bar\Delta^{++}\Delta^{++}$ & $\bar\Delta^{+}\Delta^{+}$ & 
$\bar\Delta^{0}\Delta^{0}$ & $\bar\Delta^{-}\Delta^{-}$ &  
$\bar\Sigma^+\Sigma^+$ & $\bar\Sigma^-\Sigma^-$ &  $\bar\Sigma^0\Sigma^0$ &  
$\bar\Xi^- \Xi^-$ &  $\bar\Xi^0 \Xi^0$ & $\bar\Omega^{-}\Omega^{-}$  \\
\hline\hline
$|\Delta_{\frac{3}{2}E}^{\bf 56}\rangle$ & 
 ${\scriptstyle 3\mu_u}$ & ${\scriptstyle 2\mu_u+\mu_d}$ &  
 ${\scriptstyle \mu_u+2\mu_d}$ & ${\scriptstyle 3\mu_d}$ & 
 ${\scriptstyle 2\mu_u+\mu_s}$ & ${\scriptstyle 2\mu_d+\mu_s}$ &
 ${\scriptstyle \mu_u+\mu_d+\mu_s}$ & ${\scriptstyle \mu_d+2\mu_s}$ & 
 ${\scriptstyle \mu_u+2\mu_s}$ &  ${\scriptstyle 3\mu_s}$ \\
\hline 
Theor. &  5.59 & 2.79 & 0 & -2.79 & 3.11 & -2.48 & 0.318 &
               -2.16 & 0.636 & -1.84 \\
\hline
Exp. & $3.7\sim 7.5$ &  &  &  &  &  &  &
                &  &  \\
\hline\hline
$|\Delta_{\frac{3}{2}F}^{{\bf 56}^\prime}\rangle$ & 
 ${\scriptstyle 3\mu_u}$ & ${\scriptstyle 2\mu_u+\mu_d}$ &  
 ${\scriptstyle \mu_u+2\mu_d}$ & ${\scriptstyle 3\mu_d}$ & 
 ${\scriptstyle 2\mu_u+\mu_s}$ & ${\scriptstyle 2\mu_d+\mu_s}$ &
 ${\scriptstyle \mu_u+\mu_d+\mu_s}$ & ${\scriptstyle \mu_d+2\mu_s}$ & 
 ${\scriptstyle \mu_u+2\mu_s}$ &  ${\scriptstyle 3\mu_s}$ \\
\hline 
Theor. &  5.59 & 2.79 & 0 & -2.79 & 3.11 & -2.48 & 0.318 &
               -2.16 & 0.636 & -1.84 \\
\hline\hline
$|\Delta_{\frac{1}{2}G}^{{\bf 70}}\rangle$ & 
 ${\mu_u}$ & $\frac{2\mu_u+\mu_d}{3}$ &  
 $\frac{\mu_u+2\mu_d}{3}$ & ${\mu_d}$ & 
 $\frac{2\mu_u+\mu_s}{3}$ & $\frac{2\mu_d+\mu_s}{3}$ &
 $\frac{\mu_u+\mu_d+\mu_s}{3}$ & $\frac{\mu_d+2\mu_s}{3}$ & 
 $\frac{\mu_u+2\mu_s}{3}$ & ${\mu_s}$ \\
\hline 
Theor. &  1.86 & 0.931 & 0 & -0.931 & 1.037 & -0.825 & 0.106 &
               -0.719 & 0.212 & -0.613 \\
\hline\hline
\end{tabular}
\caption{Magnetic moment of $\Delta$ decouplets: 
Unit is the nuclear magneton $\mu_N$.
%$\mu_d=-\mu_u/2$ is assumed. 
%The $\mu_u=1.862 \mu_N$ and $\mu_s=-0.613 \mu_N$ are used. 
}
\label{tab9}
\end{table}

\begin{table}
\begin{center}
\begin{tabular}{|l|c|}
\hline\hline
   & $\bar\Lambda^-\Lambda^-$ \\
\hline
$|\Lambda_{\frac{1}{2}G}^{\bf 70}\rangle$ &  $\frac{\mu_u+\mu_d+\mu_s}{3}$ \\
\hline
Theor. & 0.106\\
\hline\hline
\end{tabular}
\end{center}
\caption{Magnetic moment of $\Lambda$ singlet}
\label{tab10}
\end{table}

Here the $m.m.$ of $N(939)$ and $\Lambda (1116)$
are used as inputs, to determine
$\mu_u=1.862 \mu_N$ and $\mu_s=-0.613 \mu_N$ ($\mu_N=e/(2m_N)$ and $\mu_d=-\mu_u/2$ is assumed).
These values correspond to $g_M/m_u=1/0.336{\rm GeV}$ and $g_M/m_s=1/0.510{\rm GeV}$. 

The $m.m.$ of baryons in ${\bf 56}$ thus calculated are 
of the same values as the predictions by NRQM.
The values of the extra ${\bf 56}^\prime$-multiplet is also
the same. The values for {\bf 70}-multiplet are given in the tables, too.

\subsection{Radiative Decays}

By taking the overlapping of 
$j^{EM}_{1\mu}(P^\prime ,P) A_\mu (q)$ of Eq.~(\ref{eqC2}), 
between the initial and final WFs, $W_{\alpha \beta \gamma}$, 
we obtain the helicity amplitudes,
\begin{eqnarray}
A_{j_z}  &=& \frac{1}{ \sqrt{2|{\mib q}|} } \bar \Phi_U ({\mib v}, j_z-1) 
    3 j^{EM}_{1\mu}({\mib P},{\mib 0}) \tilde\varepsilon_\mu^{(-1)} (q) \Phi({\mib 0}, j_z),
\label{R1}
\end{eqnarray}
where we take the rest frame of initial baryon, and the photon is 
emitted to $- \hat{\mib z}$ direction
with momentum ${\mib q}$. Its helicity is taken to be -1.  
Correspondingly the baryon spin changes from $j_z$ to $j_z-1$.  
The $A_{\frac{3}{2}}$ and $A_{\frac{1}{2}}$ appear for the initial baryons with spin $3/2$, while does
only $A_{\frac{1}{2}}$ for the spin $1/2$ baryons. ($A_{-\frac{3}{2}}$ and $A_{-\frac{1}{2}}$ are obtained from
$A_{\frac{3}{2}}$ and $A_{\frac{1}{2}}$,  respectively, by operating Parity and $\pi$-rotation around $y$-axis.) 

The decay widths of the baryon with spin $3/2$ and $1/2$, denoted respectively as 
$\Gamma_{\frac{3}{2}}$ and  $\Gamma_{\frac{1}{2}}$, 
are given, in terms of $A_{j_z}$, by
\begin{eqnarray} 
\Gamma_{\frac{3}{2}} &=& \frac{|{\mib q}|^2}{2\pi}\frac{M}{M_R} \left[ |A_{\frac{3}{2}}|^2 + |A_{\frac{1}{2}}|^2 \right] , \ \ \ 
\Gamma_{\frac{1}{2}} = \frac{|{\mib q}|^2}{\pi}\frac{M}{M_R} \left[ \ \ \ \ \  |A_{\frac{1}{2}}|^2 \ \ \  \right] \ ,
\label{R4}
\end{eqnarray}
where $M_R(M)$ is the mass of initial(final) baryon.

%Among the experimentally observed baryons, the following ones have rather low masses, and may 
%be relativisitically $S$-wave chiral states: 
%$\Delta^{+\hspace{-0.25cm}\bigcirc} (1232)$, $\Delta^{+\hspace{-0.25cm}\bigcirc}  (1600)$, 
%$N^{+\hspace{-0.25cm}\bigcirc} (1440)$, $\Lambda^{+\hspace{-0.25cm}\bigcirc} (1600)$,
%$\Sigma^{+\hspace{-0.25cm}\bigcirc} (1660)$, 
%$\Lambda^{-\hspace{-0.25cm}\bigcirc} (1405)$, $N_{\frac{3}{2}}^{-\hspace{-0.25cm}\bigcirc} (1520)$, 
%$\Lambda_{\frac{3}{2}}^{-\hspace{-0.25cm}\bigcirc} (1520)$
%and $\Sigma_{\frac{3}{2}}^{-\hspace{-0.25cm}\bigcirc} (1670)$.
%We consider the radiative decays of these baryons.
%The decay widths (or helicity amplitudes) are
%observed except for $\Lambda^{+\hspace{-0.25cm}\bigcirc} (1600)$,
%$\Sigma^{+\hspace{-0.25cm}\bigcirc} (1660)$ and $\Sigma_{\frac{3}{2}}^{-\hspace{-0.25cm}\bigcirc} (1670)$.

Here we focus our interests on the decays of $\Delta^{+\hspace{-0.25cm}\bigcirc} (1232)$ and 
the candidates of chiralons,  
$\Delta^{+\hspace{-0.25cm}\bigcirc}  (1600)$, 
$N^{+\hspace{-0.25cm}\bigcirc} (1440)$ and $\Lambda^{+\hspace{-0.25cm}\bigcirc} (1600)$,
considering the possible mixing of 
their WFs defined in Table \ref{tabZ}.

\begin{table}
\begin{center}
\begin{tabular}{l|l}
  &  $|\  F\ \rho\  \sigma   \ \rangle $\\
\hline
$| N(939),\Lambda (1116),\Sigma (1192)\rangle$
  & $c_1 | N(F^{+\hspace{-0.25cm}\bigcirc} )\rangle - s_1 | N(E^{+\hspace{-0.25cm}\bigcirc} )\rangle$  \\
$| N(1440)\rangle$
  & $\left( -s"_1 | N(F^{+\hspace{-0.25cm}\bigcirc} )\rangle - c"_1 
| N(E^{+\hspace{-0.25cm}\bigcirc})\right) c_2 + s_2| N(G^{+\hspace{-0.25cm}\bigcirc}) \rangle$  \\
$| \Delta (1232) \rangle$
  & $c^\prime_1 | \Delta (F^{+\hspace{-0.25cm}\bigcirc} )\rangle - s^\prime_1 | \Delta (E^{+\hspace{-0.25cm}\bigcirc} )\rangle$  \\
$| \Delta (1600) \rangle$
  & $-s^\prime_1 | \Delta (F^{+\hspace{-0.25cm}\bigcirc} )\rangle - c^\prime_1 | \Delta (E^{+\hspace{-0.25cm}\bigcirc} )\rangle$  \\
$| \Lambda (1405)\rangle$
  & $\left[ \left( -s"_1 | N(F^{-\hspace{-0.25cm}\bigcirc} )\rangle - c"_1 
| N(E^{-\hspace{-0.25cm}\bigcirc})\right) (-s_2) + c_2| N(G^{-\hspace{-0.25cm}\bigcirc}) \right] s_3
  + c_3  | \Lambda (G^{-\hspace{-0.25cm}\bigcirc})      \rangle$  \\
\hline
\end{tabular}
\end{center}
\caption{Form of WF of the relevant baryons: $c_1$ ($s_1$) etc. are abbreviations of 
$cos\theta_1$ ($sin\theta_1$), and thus, $c_1^2+s_1^2=1$.}
\label{tabZ}
\end{table} 

In order to keep the successful values of $m.m.$ in NRQM, 
we consider only the mixing between ${\bf 56}_E$ and ${\bf 56}_F$ for 
$N(939)$-octet, neglecting the mixing with ${\bf 70}_G$. 
For $N(1440)$ the mixing with ${\bf 70}_G$ is also considered. 
For $\Lambda (1405)$ we generally consider the mixing among four $\Lambda$s, 3 octet $\Lambda$
(in {\bf 56}, ${\bf 56}^\prime$ and {\bf 70}) and one singlet (in {\bf 70}).
The radiative decay amplitudes obtained by using these WFs 
are given in Table \ref{tabA}.

\begin{table}
\begin{center}
\begin{tabular}{ll|l}
\hline
Process & $\lambda$ & \ \ \ \ \ \ \ \ \ \ \ \ $A_\lambda$\ \ \ \ \  
          (in unit of $\frac{3eg_M}{2m_u}\sqrt{|{\mib q}|}$ )  \\ 
\hline
$\Delta (1232)\rightarrow N\gamma $  & $\frac{3}{2}$   
  &  $-\frac{\sqrt{2}}{3\sqrt{3}} \left[ (s'_1s_1+c'_1c_1) (ch^3\theta +ch^2\theta sh\theta ) 
        + (-\frac{4}{3}c'_1c_1+\frac{2}{\sqrt{3}}(c'_1s_1+s'_1c_1) ) ch^2\theta sh\theta   \right]$ \\
                                     &  $\frac{1}{2}$
  &  $-\frac{\sqrt{2}}{9}  (s'_1s_1+c'_1c_1) (ch^3\theta +ch^2\theta sh\theta ) $ \\
\hline
$\Delta (1600)\rightarrow N\gamma $  & $\frac{3}{2}$   
  &  $-\frac{\sqrt{2}}{3\sqrt{3}} \left[ (c'_1s_1-s'_1c_1) (ch^3\theta +ch^2\theta sh\theta ) 
        + (\frac{4}{3}s'_1c_1+\frac{2}{\sqrt{3}}(c'_1c_1-s'_1s_1) ) ch^2\theta sh\theta   \right]$ \\
                                    &  $\frac{1}{2}$
  &  $-\frac{\sqrt{2}}{9}  (c'_1s_1-s'_1c_1) (ch^3\theta +ch^2\theta sh\theta ) $ \\
\hline
$p (1440)\rightarrow  p \gamma $   &  $\frac{1}{2}$ 
  &  $ \frac{1}{3}  (c"_1s_1-s"_1c_1) c_2 (ch^3\theta + ch^2\theta sh\theta )
        + \frac{1}{3\sqrt{3}}  \left( - s_1 - \frac{1}{\sqrt{3}} c_1 \right) s_2 ch^2\theta sh\theta $ \\
$n (1440)\rightarrow  n \gamma $  &  $\frac{1}{2}$  
  &  $ -\frac{2}{9}  (c"_1s_1-s"_1c_1) c_2 (ch^3\theta + ch^2\theta sh\theta )
       - \frac{1}{3\sqrt{3}}  \left( - s_1 - \frac{1}{\sqrt{3}} c_1 \right) s_2 ch^2\theta sh\theta $ \\
\hline
$\Lambda (1405) \rightarrow  \Sigma^0 \gamma $  &  $\frac{1}{2}$ 
  &  $\frac{1}{18}\left[ 2 (-c"_1c_1+s"_1s_1-\frac{2}{\sqrt{3}}s"_1c_1 )
                 (-s_2) s_3 (ch^3\theta + ch^2\theta sh\theta ) \right.$\\
  &  &  $\ \ \ \ +\left( - s_1 - \frac{1}{\sqrt{3}} c_1 \right) 
         \left. \left\{  c_2 s_3 (ch^3\theta - ch^2\theta sh\theta )
              -3  c_3 (ch^3\theta + 2 ch^2\theta sh\theta )\right\}  \right] $ \\
$\Lambda (1405) \rightarrow  \Lambda \gamma $  &  $\frac{1}{2}$  
  &  $ \frac{1}{18\sqrt 3} \left[   -\frac{2m_u}{m_s}  (-c"_1c_1+s"_1s_1-\frac{2}{\sqrt{3}}s"_1c_1 )
  (-s_2) s_3 (ch^3\theta + ch^2\theta sh\theta ) \right. $ \\
 &  &  $\ \ \ \ \ \ \ \ \ \ +\left( - s_1 - \frac{1}{\sqrt{3}} c_1 \right)
  \left\{ c_2 s_3 \left(  ( 1-\frac{2m_u}{m_s})ch^3\theta + \sqrt{2}ch^2\theta sh\theta \right) \right.$\\
 &  &  $\ \ \ \ \ \ \ \ \ \ \ \ \ \ \ \ \ \ \ \ \ \ \ 
        \ \ \ \ \ \ \ \ \ \ \ \left. \left.  -c_3   \left( 1+\frac{2m_u}{m_s} \right)  
       (ch^3\theta + 2 ch^2\theta sh\theta ) \right\} \right] $ \\
\hline
\end{tabular}
\end{center}
\caption{Formula of helicity amplitudes in unit of $\frac{3eg_M}{2m_u}\sqrt{|{\mib q}|}$.}
\label{tabA}
\end{table} 

In the ideal case, case A, the WFs, corresponding to
$c_1=c_1^\prime =c"_1$ and $s_1=s_1^\prime =s"_1$, 
are orthogonal to each other. 
In case B considering the possible mixing effects with radially excited states,
we relax these relations. 

({\it $\Delta (1232)\rightarrow N\gamma$ and $\Delta (1600)\rightarrow N\gamma$})\ \ \ 
There is a longstanding problem in the radiative decay  $\Delta (1232)\rightarrow N\gamma$.
The decay width predicted by NRQM is small, and inconsistent with the experimental data.

In case A, $A_{\frac{3}{2}}= -\frac{\sqrt{2}}{3\sqrt{3}} 
[ (ch^3\theta +ch^2\theta sh\theta (1-\frac{4}{3}c_1^2+\frac{4}{\sqrt{3}} c_1s_1 ) ]$
and $A_{\frac{1}{2}}= -\frac{\sqrt{2}}{9} (ch^3\theta +ch^2\theta sh\theta )$
(in unit of $\frac{3eg_M}{2m_u}\sqrt{|{\mib q}|}$).
In non-relativistic limit, $ch\theta=1$ and $sh\theta =0$, and the helicity amplitudes take simple forms,
$A_{\frac{3}{2}}^{NR}=-\frac{\sqrt{2}}{3\sqrt{3}} $ and $A_{\frac{1}{2}}^{NR}=-\frac{\sqrt{2}}{9} $.
In the actual case of $\Delta (1232)\rightarrow N(939)\gamma$,
$sh\theta = 0.1366$, which represents the strength of relativistic recoil effect. 
In $A_{\frac{3}{2}}$ the second term proportional to $ch^2\theta sh\theta$  
 is dependent upon the mixing coefficients $c_1$ and $s_1$. 
The factor $(1-\frac{4}{3}c_1^2+\frac{4}{\sqrt{3}} c_1s_1 )$ 
can take the value from $-1$ to $\frac{5}{3}$, corresponding to 
 $(c_1,s_1)=(\frac{\sqrt 3}{2}, -\frac{1}{2})$ and $( -\frac{1}{2}, -\frac{\sqrt 3}{2})$, respectively.  
In the latter case,  the numerical values of $A_j$ are improved as
\begin{eqnarray}
A_{\frac{3}{2}}^{NR}=-0.187 &\rightarrow& A_{\frac{3}{2}}=-0.236,\ \ \ 
%A^{exp}_{\frac{3}{2}}=-0.255\pm 0.008 
A_{\frac{1}{2}}^{NR}=-0.108 \rightarrow A_{\frac{1}{2}}=-0.126\ \ \ 
%A^{exp}_{\frac{1}{2}}=-0.135\pm 0.006\ ,
\label{eq111}
\end{eqnarray}
which are almost consistent with experimental data shown in Table \ref{tabB}.
Correspondingly the decay width is improved as $\Gamma^{NR}=379$keV $\rightarrow$ $\Gamma_{theor}=581$keV,
which is almost consistent with experimental value $\Gamma_{exp}=624\sim 720$keV. 
By changing the mixing angle, $\Gamma_{theor}$ can take the value from 353keV to 581keV.
Thus, the problem of the $\Delta \rightarrow N \gamma$ width may be solved by considering the relativistic effect.
It is notable that,
in case of maximum $\Gamma_{theor}$ with $\frac{4}{3}s_1c_1-\frac{2}{\sqrt 3}(c_1^2-s_1^2)=0$, 
the $\Delta (1600)$ decay amplitudes vanishes consistently with experiment.
(In case B we show the result obtained by using $A_{\frac{1}{2}}$ of $\Delta (1600)$ decay as input,
which gives $(c'_1s_1-s'_1c_1)=0.11\pm 0.09$.)
%This gives further support for the validity of our relativistic scheme. 

({\it $N (1440)\rightarrow N\gamma$})\ \ \ As shown in Table \ref{tabA} the effect of ${\bf 70}_G$-mixing 
represented by $s_2$ gives the ratio $n\gamma /p\gamma =-1$, which seems to be inconsistent with 
the experimental value $\sim -\frac{2}{3}$. 
So we take $s_2=0$. In case A, $c"_1s_1-s"_1c_1=0$, and thus, both $A_{\frac{3}{2}}$ and $A_{\frac{1}{2}}$
vanish, being inconsistent with experimental data. In case B we take the value  $c"_1s_1-s"_1c_1=-0.1728$,
which reproduces the data well.  

({\it $\Lambda (1405)\rightarrow \Lambda\gamma ,\ \Sigma^0\gamma$})\ \ \ In case A of
maximum $\Delta \rightarrow N\gamma$, $(-c_1^2+s_1^2-\frac{2}{\sqrt{3}}s_1c_1 )=0$, and thus,
the effect on $A_{\frac{1}{2}}$  from the mixing with ${\bf 56}_E$ and ${\bf 56}_F$ vanishes. 
So, we consider only the mixing 
between octet $\Lambda_{8,G}$ and singlet $\Lambda_{1,G}$ in ${\bf 70}_G$, 
taking $c_2=1$.
In case of $\Lambda (1405)$ being purely $\Lambda_{1,G}$ ($c_3=1$), 
$(\Gamma_{\Lambda\gamma}, \Gamma_{\Sigma^0\gamma})=(98,195)$keV, 
while in case of purely $\Lambda_{8,G}$ ($s_3=1$),  (0.3,17)keV. Both cases are inconsistent with the experiment.
By taking $(c_3,s_3)=(0.478,0.878)$, the experimental $\Gamma_{\Lambda\gamma}$ is reproduced,
and in this case  $\Gamma_{\Sigma^0\gamma}$ is predicted with $15.9$keV, which seems to be consistent
with the present experimental values. 

\begin{table}
\begin{center}
\begin{tabular}{ll|ll|l||ll|l}
\hline
Process & $\lambda$ & $A_{\lambda ,\ case\ A}^{\rm theor}$ &   $A_{\lambda ,\ case\ B}^{\rm theor}$ &
              $A_\lambda^{\rm exp}$(GeV$^{-\frac{1}{2}}$)
 & $\Gamma^{tot}_{\rm case A}$ &  $\Gamma^{tot}_{\rm case B}$  & $\Gamma^{tot}_{\rm exp}$(keV) \\ 
\hline
$\Delta (1232)\rightarrow N\gamma $ 
   & $\frac{3}{2}$ & -0.236 & $>-0.236$ & -0.255$\pm$0.008 & 581 & $<581$ &  624$\sim$720 \\ 
   & $\frac{1}{2}$ & -0.126 & $>-0.126$ & -0.135$\pm$0.006 &     &        &   \\ 
\hline
$\Delta (1600)\rightarrow N\gamma $ 
   & $\frac{3}{2}$  & 0 &  -0.040$\pm$0.035 & -0.009$\pm$0.021 & 0 & 1$\sim$192 & 3.5$\sim$70\\  
   &  $\frac{1}{2}$ & 0 & \underline{-0.023$\pm$0.020} & -0.023$\pm$0.020 & & & \\  
\hline
$p (1440)\rightarrow  p \gamma $ 
   &  $\frac{1}{2}$ & 0 & \underline{-0.065} & -0.065$\pm$0.004 & 0  & \underline{150} & 150$\pm$18 \\  
$n (1440)\rightarrow  n \gamma $ 
   &  $\frac{1}{2}$ & 0 & 0.044 & 0.040$\pm$0.010 & 0 & 69 & 57$\pm$28 \\   
\hline
$\Lambda (1405) \rightarrow  \Lambda \gamma $ 
   &  $\frac{1}{2}$ & \underline{-0.040} & able & 0.040$\pm$0.005 & \underline{27}  & able &  27$\pm$8 \\
$\Lambda (1405) \rightarrow  \Sigma^0 \gamma $ 
   &  $\frac{1}{2}$ & -0.039 & to fit 
     & $\stackrel{0.031\pm 0.005}{\scriptscriptstyle 0.047\pm 0.006}$ & 16 & to fit 
     & $\stackrel{10\pm 4}{\scriptscriptstyle 23\pm 7}$ \\ 
\hline
\end{tabular}
\end{center}
\caption{Predicted values of helicity amplitudes (in unit of GeV$^{-\frac{1}{2}}$) 
in comparison with experiments. 
Decay widths are also given (in unit of keV).}
\label{tabB}
\end{table}

\section{Concluding Remarks}

We have investigated the electromagnetic properties of ground state baryons 
in the covariant level-classification scheme.
The longstanding puzzle on reproducing the large experimental 
$\Delta (1232) \rightarrow N \gamma$ amplitudes 
is solved by considering the mixing with relativisitic ${\bf 56}^\prime_F$ states. 
This mixing explains simultaneously 
the vanishing $\Delta (1600)$ decay amplitudes.
This fact strongly supports the validity of our covariant level classification scheme. 
This scheme predicts the existence of many other chiral states not describable in NRQM.   
The search for them is urgently required.

\end{document}